# Approximation of Statistical Analysis and Estimation by Morphological Adaptation in a Model of Slime Mould

Jeff Jones and Andrew Adamatzky

Centre for Unconventional Computing
University of the West of England
Coldharbour Lane
Bristol, BS16 1QY, UK.
`jeff.jones@uwe.ac.uk, andrew.adamatzky@uwe.ac.uk`

**Abstract.** True slime mould *Physarum polycephalum* approximates a range of complex computations via growth and adaptation of its protoplasmic transport network, stimulating a large body of recent research into how such a simple organism can perform such complex feats. The properties of networks constructed by slime mould are known to be influenced by the local distribution of stimuli within its environment. But can the morphological adaptation of slime mould yield any information about the *global* statistical properties of its environment? We explore this possibility using a particle based model of slime mould. We demonstrate how morphological adaptation in blobs of virtual slime mould may be used as a simple computational mechanism that can coarsely approximate statistical analysis, estimation and tracking. Preliminary results include the approximation of the geometric centroid of 2D shapes, approximation of arithmetic mean from spatially represented sorted and unsorted data distributions, and the estimation and dynamical tracking of moving object position in the presence of noise contaminated input stimuli. The results suggest that it is possible to utilise collectives of very simple components with limited individual computational ability (for example swarms of simple robotic devices) to extract statistical features from complex datasets by means of material adaptation and sensorial fusion.

## 1 Introduction - Morphological Computation

Morphological computation seeks to exploit the embodied properties of physical, living, and synthetic systems for useful computation [28]. Influences on morphological computation can be found in the natural world where apparently simple, low-level mechanisms generate complex patterning seen at very different scales including car traffic dynamics [10], human walking patterns [11], flocking and schooling [30], collective insect movement [5], and bacterial patterning [24]. In all these examples there is a population of entities in space, coupled by sensory information about their environment. The collective morphology of the group is generated from the local interactions and movement of individual members of the population. These interactions generate complex, self-organised and emergent behaviour. The resulting morphological patterns of the population show adaptation to the environment, for example avoidance of obstacles, detection of prey, and avoidance of predators.

The complex adaptation in seen biological systems suggests a literal form of morphological computation, driven by dynamic patterning in a changing environment. An ideal candidate for a morphological computation medium would be a material capable of the complex sensory integration, movement and adaptation of a living organism, yet composed of relatively simple components that are amenable to understanding and control. The giant single-celled amoeboid organism *Physarum polycephalum* meets both criteria. During the plasmodium stage of its life cycle it adapts its body plan in response to a range of environmental stimuli (nutrient attractants, repellents, predators) during its growth, foraging and feeding. The plasmodium is composed of a transport network of protoplasmic tubes which spontaneously exhibit contractile activity which is used to distribute nutrients throughout its body plan.

Plasmodium of slime mould is amorphous in shape and ranges from the microscopic scale to up to many square metres in size. It is a giant syncytium formed by repeated nuclear division, comprised of a sponge-like actomyosin complex co-occurring in two physical phases. The gel phase is a dense matrix subject to spontaneous contraction and relaxation, under the influence of changing concentrations of intracellular chemicals. The protoplasmic sol phase is transported through the plasmodium by the force generated by the oscillatory contractions within the gel matrix. Protoplasmic flux, and thus the behaviour of the organism, is affected by local changes in temperature, space availability, chemoattractant stimuli and illumination [6], [22], [25], [27]. The *Physarum* plasmodium can thus be regarded as a complex functional material capable of both sensory and motor behaviour. Indeed *Physarum* has been described as a membrane bound reaction-diffusion system in reference to both the complex interactions within the plasmodium and the rich computational potential afforded by its material properties [3]. The study of the computational potential of the *Physarum* plasmodium was initiated by Nakagaki et al. [26] who found that the plasmodium could solve simple maze puzzles. This research has been extended and the plasmodium has demonstrated its performance in, for example, path planning and plane division problems [33], spanning trees and proximity graphs [2], [1], simple memory effects [31], the implementation of individual logic gates [35] and *Physarum* inspired models of simple adding circuits [16].

## 2 Basic Method and Overview of Paper

Although slime mould has desirable computational properties, it also has limitations due to the fact that it is a living organism. Although relatively simple and inexpensive to culture, slime mould is also slow (compared to classical computing substrates) and must be maintained within strict environmental conditions. It may also be relatively unpredictable in its behaviour. Although unpredictability is useful in wild conditions, it can be a hindrance when repeatable measures of its performance are required. We therefore require a synthetic model of slime mould.

We used the multi-agent approach introduced in [13]. Agents sense the concentration of a hypothetical 'chemical' in a 2D lattice, orient themselves towards the locally strongest source and deposit the same chemical during forward movement. The agent population spontaneously forms emergent transport networks which undergo complex evolution, exhibiting minimisation and cohesion effects under a range of sensory parameter settings. The dynamical network patterns were found to reproduce a wide range of Turing-type reaction-diffusion patterning [12]. External stimuli by nutrients and repellents are represented by projecting positively weighted and negatively weighted values respectively into the lattice and the network evolution is constrained by the distribution of nutrients and repellents. Network evolution is affected by nutrient distribution, nutrient concentration and repellent placement. The networks formed by the model reproduced the connectivity of slime mould by approximating networks in the Toussaint hierarchy of proximity graphs [14], as originally demonstrated in [1]. Using a combination of attractant and repellent stimuli the method is also capable of approximating the Convex Hull, Concave Hull and Voronoi diagram [19]. By using a feedback process to dynamically control the concentration of attractant stimuli the networks were able to approximate simple instances of the Travelling Salesman Problem (TSP) [15].

Much of the previous research using this model has concentrated on its network minimisation behaviours, however the bulk properties of larger collectives has also been studied. In [36] larger population sizes were found to reproduce the oscillatory patterns and transitions seen in *Physarum* and these aggregates were found to exhibit controllable collective amoeboid movement [17] and distributed conveyor transport [37]. These larger aggregates have also been found to be capable of material computation: In [20] a simple unguided method based on material shrinkage was shown to have good performance on instances of the TSP.

In this paper we continue to explore the computing properties of relatively large aggregate blobs of the multi-agent collectives by examining whether it is possible to extract global statistical properties of datasets via indirect means using only local interactions within the multi-agent population. In section 3 we take the phenomenon of sclerotium formation in *Physarum* as an inspiration to develop a spatially represented unconventional computing mechanism based upon material cohesion, shrinkage and adaptation. Using this mechanism we extract geometric properties, specifically the centroid, of 2D datasets via shrinkage and adaptation of the agent population. In Section 4 we examine the performance of the shrinkage method on spatial arrangements of 1D numerical datasets in extracting the arithmetic mean. The ability of the collective to cope with dynamically changing stimuli is assessed in Section 5 in which we attempt to use the blob to track the position of a simulated mobile target in the presence of noise-free and noise-contaminated stimuli, and using different stimulus types. The mechanisms of cohesion, shrinkage and shape adaptation by which the material computation is performed is described in Section 6, including some limitations and drawbacks of the approach. We conclude in Section 7 by examining the properties that enable these bulk collectives to extract global information and suggest further potential applications in statistics and robotics. Due to the dynamical nature of the material adaptation the reader is encouraged to refer to the supplementary video recordings at http://uncomp.uwe.ac.uk/jeff/statistics_estimation.htm. A full description of the model and its parameters for these experiments is given in the appendix.

## 3 Sclerotinisation as an Inspiration for Centroid Computation

The sclerotium stage is a part of the life cycle of *Physarum*, whose entry is provoked by adverse environmental conditions, particularly by a gradual reduction in humidity. In prolonged dry conditions the mass of plasmodium aggregates together, abandoning its protoplasmic tube network to form a compact, typically circular or elliptical, toughened mass [21]. Sclerotinisation protects the organism from environmental damage and the slime mould can survive for many months — or even years — in this dormant stage, re-entering the plasmodium stage when moist conditions return. Biologically, the sclerotium stage may be interpreted as a primitive survival strategy and it has been interpreted computationally as a biological equivalent of freezing or halting a computation [4] in spatially represented biological computing schemes.

Does the position of the sclerotium in *Physarum* exhibit any regular properties? And, if so, can the phenomenon of sclerotinisation in *Physarum* serve as an inspiration for a mechanism of geometric material computation? To assess sclerotium position we patterned a set of oat flakes in the arrangement of 23 most populous cities of the Iberian peninsula and inoculated the plasmodium in the Madrid region (Fig. 1a). The plasmodium colonised all the oat flakes within 58 hours and sclerotium formation was then initiated by gradual desiccation. An example of a sclerotium is shown in Fig. 1b. The positions of sclerotium formation over 20 experiments are superimposed in Fig. 1h. The sclerotia adopt a variety of different area coverage patterns in approximately circular or elliptical shapes. Dense areas, representing more frequent sclerotium positions, can be seen in the Madrid region (occurring in 50% of experiments) and also in more southerly regions in 80% of experiments (Fig. 1h). These patterns, although *suggestive* of some regularity in aggregation, obviously show too much variability to claim that the plasmodium approximates any computation in the formation of the sclerotium position. Possible reasons why regular sclerotium position was not formed include the variations in the current active zone of the plasmodium immediately before sclerotinisation, variations in protoplasmic transport, and the layout of the tube network and slime capsule before sclerotium formation (*Physarum* has recently been shown to be sensitive to its previous locations, as recorded by its slime capsule [29]). It is also not fully understood how environmental factors affect sclerotium formation. Do influences such as the drying agar substrate influence the plasmodium? For example, if outer regions of the agar dry more quickly and harden before inner regions, the differences in substrate hardness may influence plasmodium movement [34].

The above factors, coupled with the innate network formation behaviour of the organism, render it infeasible to uniformly shrink the organism under controllable conditions. The process of sclerotium formation must therefore be interpreted as an inspiration for spatially represented computational mechanisms, rather than as a computation in itself. To investigate computation by shrinkage and aggregation mechanisms we use the particle model which can uniformly respond and adapt to changes in synthetically imposed environmental conditions. We must approximate the environmental conditions which provoke sclerotium formation. We do this by simulating the removal of all stimuli, thus allowing the virtual plasmodium to adapt and shrink in time. The coarse results presented in Fig. 1h suggest that the plasmodium may approximate the centroid of the distribution of nutrients. The geometric centroid is a weighted mean of all the X and Y co-ordinates of a shape. For a two-dimensional shape with uniform thickness the centroid can be considered as the centre of mass of the shape and, for certain complex shapes, the centre of mass may lie outside the shape itself. Can the virtual plasmodium shed any light on any computation — or otherwise — by the *Physarum* plasmodium?

As an initial assessment we patterned the virtual material into the entire shape of the Iberian peninsula Fig. 1i (the white area represents the initial pattern of the blob of virtual material). The coverage of the entire shape simulates a completely uniform transport network and eliminates any bias by a pre-selected network configuration. The blob is held in place for a short period (50 scheduler steps) by projecting attractant in the original pattern, before all attractant is removed. The blob then adapts its shape, using its innate relaxation behaviour to adopt a minimal circular shape. The blob is reduced in size by randomly removing particles from the mass ($p = 0.0005$ removal, per particle, per scheduler step) and the initial shape of the collective adapts and shrinks to a small mass (Fig. 1c-g) mimicking sclerotium formation. The final centre position of the blob was recorded over ten runs and is shown in Fig. 1i as an aggregation of blue dots, compared to the exact centroid position indicated by the red cross. Surprisingly, despite the stochastic influences on the model, the mean absolute error in position of the blob 'sclerotium' compared to the exact centroid position was only 3.39 pixels with a standard deviation of 1.35. Although it must be emphasised that the shrinkage and adaptation of the virtual blob is a very simple approximation of sclerotium formation this does suggest, in a wider com-putational sense, that the physical adaptation over time (via shape minimisation and shrinkage) may abstract some computationally (and perhaps biologically...) useful information about the original configuration. These preliminary results led us to further investigate methods and results in the approximation of statistical properties of complex datasets by morphological adaptation.

To assess different morphological adaptation methods and to see how well the adaptation approximates the centroid we initiated a large mass of virtual material in the pattern of a number of shapes. The shapes selected have different properties, such as solid, containing holes, concave, and convex. The material was held in the initial pattern by projecting attractants into the lattice corresponding to the original pattern for 50 scheduler steps. The centroid of each of the original patterns was computed conventionally by the mean value of all points within the pattern (for example Fig. 2a, circled). Since the particle population was initially configured as the original pattern the centroid of the population obviously initially matched the centroid of the original. The attractant stimuli was then completely removed from the lattice and the material underwent morphological adaptation via its emergent relaxation behaviour. Two morphological adaptation mechanisms were investigated. In the first method shape adaptation was separate from shrinkage and the blob was allowed to adapt its shape whilst retaining the same number of particles. Shrinkage was delayed for 5000 scheduler steps and only shape adaptation was initially used. After the material has recovered an approximately circular shape (at 5000 scheduler steps) the population was then reduced in size by randomly removing particles from the blob (at probability $p = 0.0005$). As particles were removed the blob automatically shrunk in size, the shrinkage of the blob allowing a visual result of the centroid position (Fig. 2a-f). In the second method the shrinkage was implemented immediately after the stimulus was removed and occurred simultaneously with the shape adaptation (Fig. 2g-l). During both methods the centroid of the virtual material is computed conventionally by averaging the co-ordinates of all particles within the blob and compared to the centroid of the original pattern and the experiments were halted when the population size became <50. The Euclidean distance between the original centroid and blob centroid (the mean absolute error, MAE) over ten runs is shown in the graphs for delayed shrinkage (Fig. 2m) and immediate shrinkage (Fig. 2n) respectively.

The results for the lizard shape indicate that as the material adapts to the removal of stimuli and the shrinkage process, it is able to approximate and maintain the same centroid position as the original shape to within - on average - two pixels accuracy. For the delayed method (MAE 3.39, $\sigma$ 1.52), the centroid is tracked most accurately during the initial adaptation phase (< 5000 steps) and the error then increases slightly during the shrinkage process (Fig. 2m). The error increases towards the end of the shrinkage process (> 12000 steps) because the reduced number of particles in the blob allows random perturbations within the collective to be amplified. For the immediate shrinkage method the error accumulates more quickly then stabilises after 6000 steps (MAE 2.09, $\sigma$ 1.36).

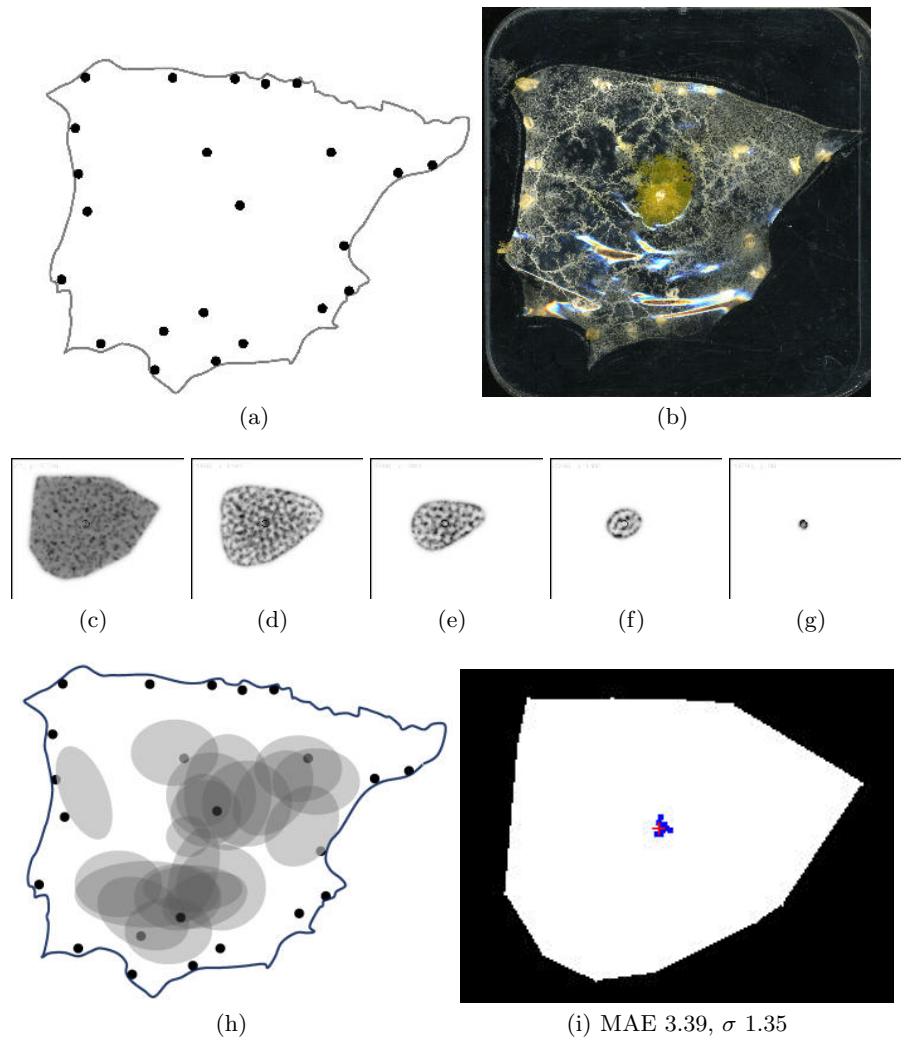

**Fig. 1.** Sclerotinisation as an approximation of the computation of centroid. a) Distribution of 23 most populous cities on the mainland Iberian peninsula, b) example showing sclerotinisation of *Physarum* and remnants of protoplasmic tubes, c-g) a shrinking blob of virtual material initialised in the Convex Hull of the cities approximates the centroid as it shrinks, h) overlay of final sclerotium positions over 20 separate experiments, i) virtual blob inoculated on the Convex Hull of the Iberian dataset showing actual centroid (red cross) and superimposed final blob position (blue squares).

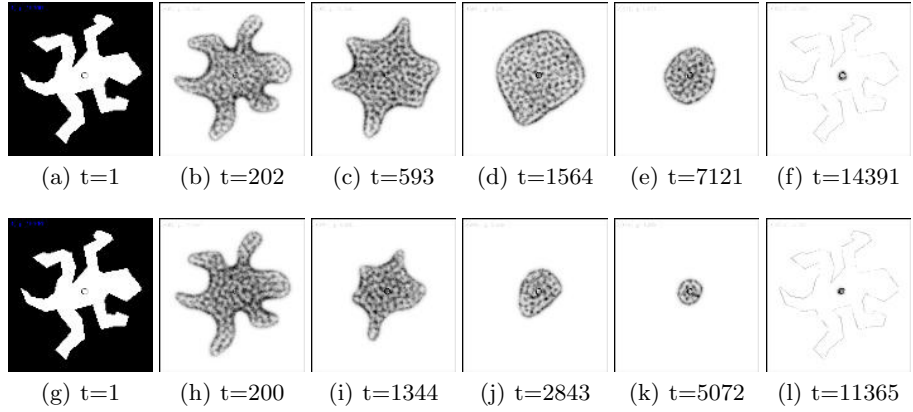

(a) t=1  (b) t=202  (c) t=593  (d) t=1564  (e) t=7121  (f) t=14391

(g) t=1  (h) t=200  (i) t=1344  (j) t=2843  (k) t=5072  (l) t=11365

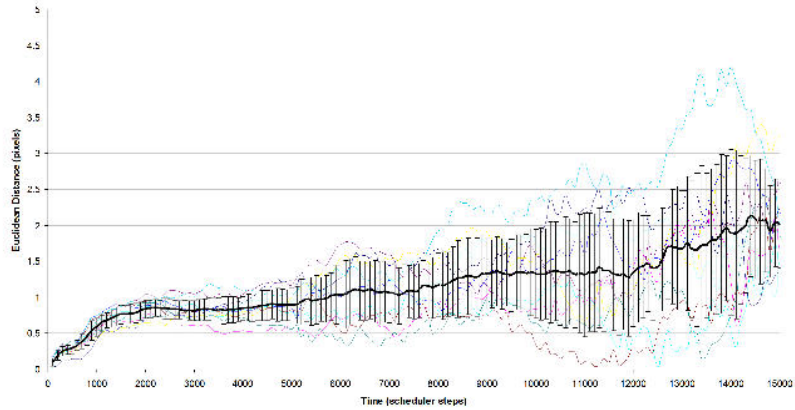

(m) delayed shrinkage, MAE 3.39, $\sigma$ 1.52

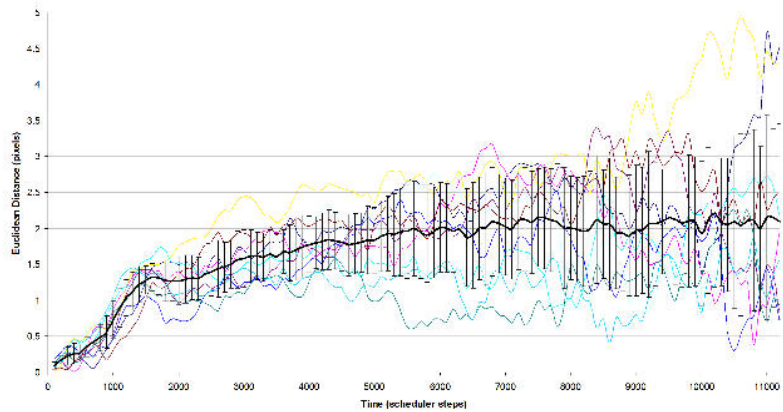

(n) immediate shrinkage, MAE 2.09, $\sigma$ 1.36

**Fig. 2.** Approximating the centroid by morphological adaptation and shrinkage. a) initial lizard shape with centroid indicated (circle), b-d) After initialisation in the original pattern the material undergoes adaptation, relaxing to approximate a circular shape, e-f) material reduces in size when particles are removed, g-l) adaptation with simultaneous shrinkage, m-n) charts plotting mean absolute error of blob centroid from original image centroid during adaptation then shrinkage (m) and adaptation with simultaneous shrinkage (n). 10 runs are shown overlaid (faint lines) with mean (thick line) and standard deviation error bars.

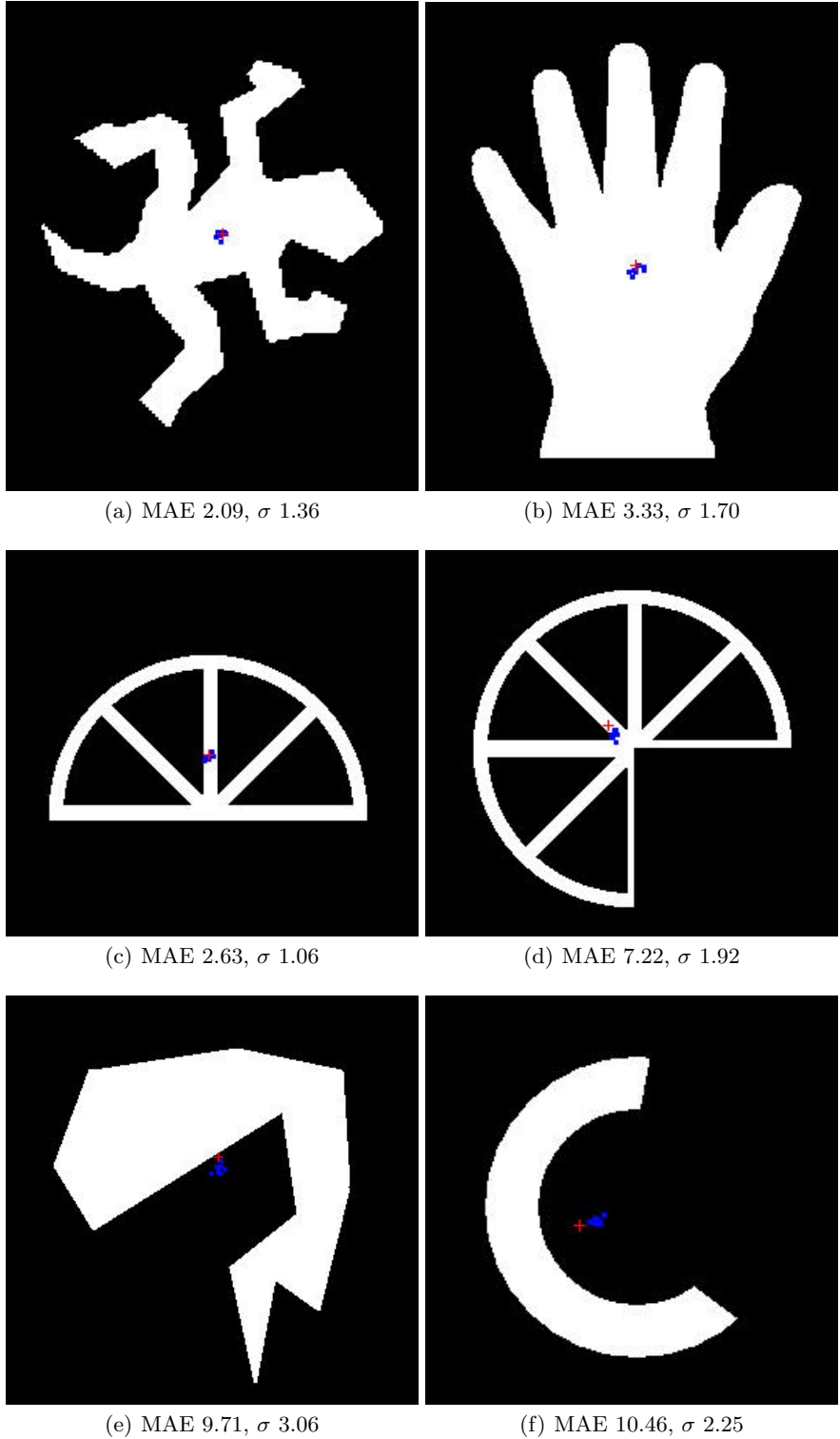

(a) MAE 2.09, $\sigma$ 1.36  (b) MAE 3.33, $\sigma$ 1.70

(c) MAE 2.63, $\sigma$ 1.06  (d) MAE 7.22, $\sigma$ 1.92

(e) MAE 9.71, $\sigma$ 3.06  (f) MAE 10.46, $\sigma$ 2.25

**Fig. 3.** Illustration of difference between blob approximation of centroid and image centroid. a-f) original image with centroid postion (red cross shape) and distribution of blob positions over 10 runs (blue dots) with mean absolute error (MAE) and standard deviation ($\sigma$) indicated in labels.

Results for the variety of shapes using the immediate shrinkage method are shown in Fig.3. Each sub-figure shows the centroid position of the original shape (marked by a red cross) and the distribution of blob centroids (the final position of the blob after adaptation and shrinkage) over ten runs, marked as a distribution blue dots. The results show better performance at tracking the centroid of convex shapes, including those with holes (Fig.3a-c). As shapes become increasingly concave, the error begins to increase. The worst performance is on shapes with strongly concave features where the centroid lies outside the boundary of the original shape (Fig.3e-f).

## 4 Approximation of Arithmetic Statistics

Instead of aggregating the geometric properties of a shape in two dimensions can we utilise the shrinkage method to summarise arithmetic properties of a spatially represented numerical dataset? To assess this possibility we randomly generated 20 numbers from the range of 0 to 100 and used these values as Y axis positions. The generated data values were sampled from a uniform distribution but had a wide range of variance across all experiments ($\sigma$ of between 542 and 1293 for the sorted data experiments and $\sigma$ between 462 and 1159 for the unsorted data experiments). X axis positions were generated using regular spacing of 20 pixels between the data points and we then connected these data points to give a shaped path on which to initialise the virtual material. The method was assessed over 50 randomly generated datasets for both unsorted lists of data (Fig. 4a,c,e,g,i,k) and for data points pre-sorted by value (Fig. 4b,d,f,h,j,l). During each run the virtual material was initially held in place by attractant projection for 20 steps of the model (Fig. 4a) and the attractant was then removed, causing the adaptive population to smooth and shrink the original shape. For unsorted data values the initial behaviour was to shrink away from sharp peaks and troughs connecting the data points (Fig. 4c) until an approximately smooth line was formed (Fig. 4e). This band of material then shrunk horizontally from each end (Fig. 4g,i). Each experiment was halted when the population size of the blob was < 50 and the final Y-axis position of the centre of the remaining population was compared to the arithmetic mean of the original data (Fig. 4k). In the case of the pre-sorted data values the smoothing of the line was much more short lived and the line began shrinking from both ends almost immediately. For the unsorted data points the mean error of the final blob position when compared to the numerically calculated arithmetic mean was 5.90 pixels ($\sigma$ 3.7) and for the pre-sorted population the mean error was 2.23 pixels ($\sigma$ 1.72). We did not find any strong correlation between the standard deviation of the randomly generated data points and performance (error) of the final position of the blob (Pearson correlation coefficient of $\rho = $ -0.07 for sorted data and $\rho = $ 0.09 for unsorted data).

How is the material shrinkage mechanism affected by skewed data distributions? To assess this we altered that random number generation procedure to generate skewed data distributions by the following method: For each of the 20 numbers generated we selected from the range of 80—100 with $p = 0.9$ and from the range 0—20 with $p = 0.1$, thus generating a list of numbers that was heavily biased towards higher numbers. We ran 25 experiments for both unsorted and sorted datasets. For unsorted skewed data (Fig. 5a,c,e,g,i, note that the Y-axis is inverted and larger values are lower) the error was 5.69 pixels ($\sigma$ 4.07), similar to the results obtained for the uniformly distributed data. For the pre-sorted skewed data, however, the error increased to 10.57 ($\sigma$ 3.04) and the final value of the blob was found to be higher than the arithmetic mean in every case (Fig. 5b,d,f,h,j). This difference from the arithmetic mean and other simple variants of the mean (such as the harmonic mean or geometric mean which are both lower than the arithmetic mean) suggests that the final blob position was influenced, or *weighted* in some way, by the greater number of larger data values. These results suggest that using morphological adaptation as a spatially implemented means of unconventional computation can — at least coarsely — approximate the arithmetic mean of a set of numbers.

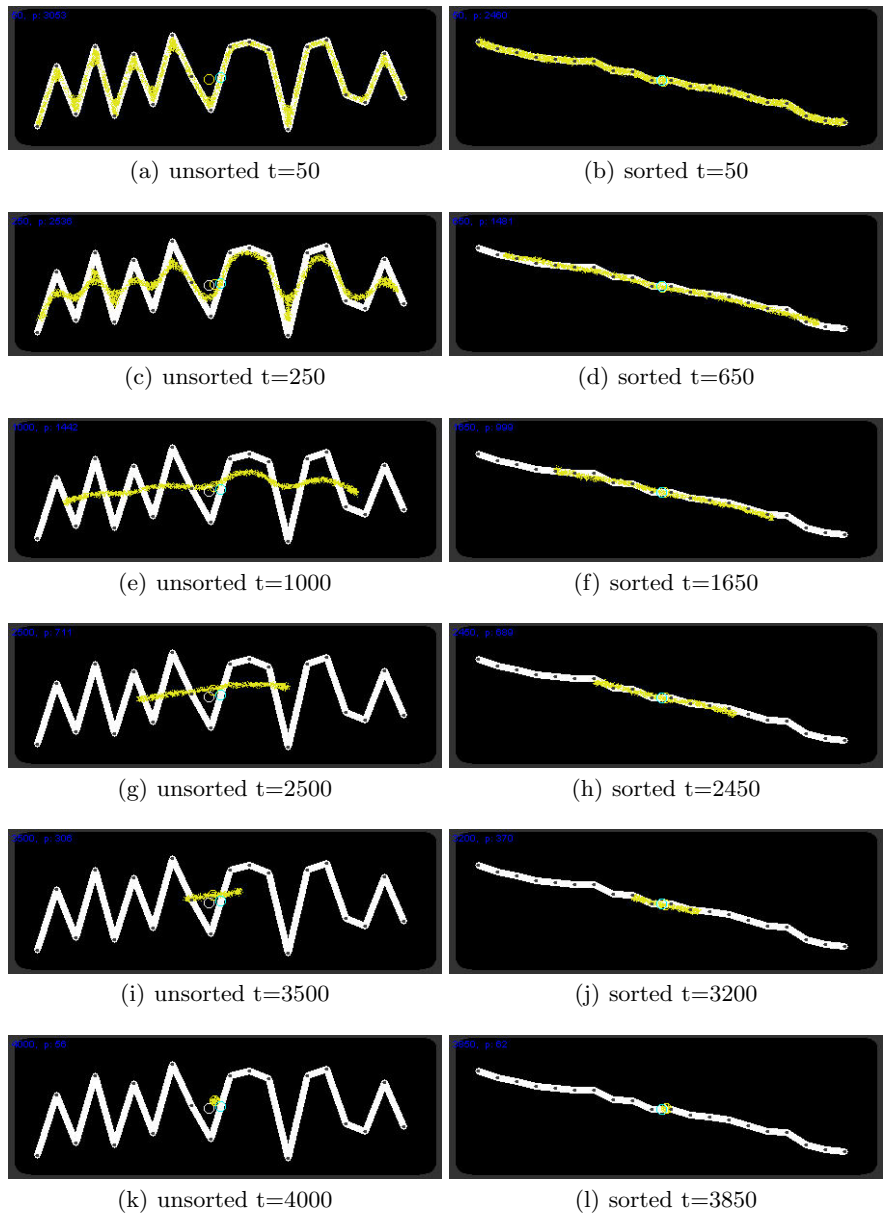

**Fig. 4.** Approximation of arithmetic mean of 1D data by morphological adaptation. Left column shows adaptation of unsorted data points, right column shows sorted data points. Individual data points indicated on inverted Y-axis by dark dots on connected line, adaptive population shown as coarse shrinking blob.

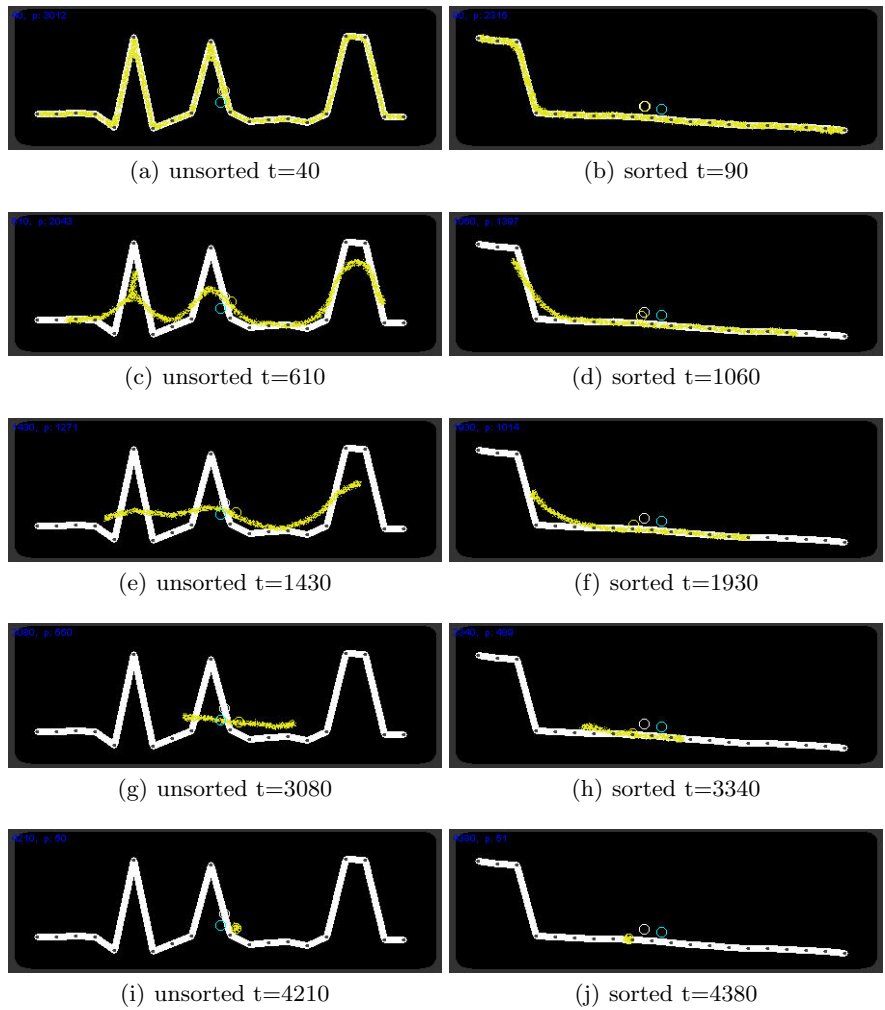

**Fig. 5.** Approximation of arithmetic mean of skewed 1D data distribution. Left column shows adaptation of unsorted data points, right column shows sorted data points. Individual data points indicated on inverted Y-axis by dark dots on connected line, adaptive population shown as coarse shrinking blob.

## 5  Spatial Approximation of Noisy Estimation Functions

Estimation functions seek to provide accurate information (output) about certain properties of a physical system using inputs which are corrupted or unreliable in some way. Estimation may be by simple statistical summation of previous inputs, or can be aided by construction of a model of the system or recursive knowledge about previous estimation errors [9]. An example application is the tracking of the actual position of a vehicle [38] given inputs (for example image snapshots or GPS co-ordinates) which are corrupted with noise, thus providing incorrect information about the position of the vehicle at any single positional update. The model of vehicle behaviour may have certain assumptions about normal vehicle movement. For example, vehicles typically move forwards or backwards and steer to the left or to the right or their current position. This simple model information suggests, for example, that sudden crab-like movement sideways without moving forwards is unusual and likely to be caused be erroneous position data.

Organisms such as *Physarum* slime mould live in an environment which is extremely noisy, from a signals perspective. Moreover *Physarum* does not possess any special senses or neural systems which are necessary in higher creatures to determine the presence of nutrients or threats. Yet *Physarum* has been shown to be capable of complex choices in terms of optimal path selection [26], nutrient quality [23] and nutrient composition [7]. The organism is able to integrate complex spatial and temporal information in a distributed manner throughout its amorphous body plan. Behavioural selection based on the sensorial fusion of disparate input stimuli is via changes in streaming and overall movement direction. Chemotaxis gradients (both attractant and repellent) coming into contact with the plasmodium affect the properties of the membrane, in terms of membrane softening or changes in local oscillation frequencies, ultimately causing changes in streaming behaviour towards attractants and away from repellents.

Can such simple material deformation mechanism be employed synthetically to provide coarse estimation functions? To assess this possibility we employed oscillatory 'blobs' of the virtual material to try to dynamically track the position of a moving stimulus whose exact position $p$ is corrupted by Gaussian noise. In the oscillatory mode of behaviour the individual particles comprising the collective may exhibit transient resistance to obstructions in movement. These regions of transient resistance interact locally with regions containing vacant spaces, resulting in interacting and competing travelling waves and emergent amoeboid movement [36], [17]. Under normal non-oscillatory conditions any particles whose movement is obstructed simply select a random new direction, resulting in smooth morphological adaptation (see the Appendix for a more complete distinction between oscillatory and non-oscillatory particle modes). The oscillatory movement conditions were used for the noisy estimation approximation because they allow faster morphological adaptation in the face of rapidly changing environmental stimuli and a blob of oscillating virtual material also exhibits directional persistence towards more recent stimuli. This directional persistence acts as a very simple approximation and model of vehicular directional persistence.

We examined noisy estimation of object tracking in 2D environments using attractant ($+ve$) stimuli, repellent ($-ve$) stimuli and a fusion of both attractant and repellent ($\pm ve$) stimuli. 2D environments were represented by placing the blob in the middle of a square lattice, tracking movement of a simulated vehicle moving on the X-Y plane. Attractant stimuli were represented by the temporary projection of point attractant at stimulus locations into the lattice for a 10 scheduler steps. Attractant stimuli near the periphery of the blob caused the blob to stream towards the stimulus. Repellent stimuli were represented by simulated illumination of the environment. The Area inside a square region surrounding the stimulus location ($50 \times 50$ pixels wide, corresponding to the approximate size of the blob area) was masked off and all regions outside this square were exposed to temporary projection (for 10 scheduler steps) of simulated illumination. Illumination was simulated by the following method: If an agent is located within an illuminated region the value of sensed chemoattractant is reduced by multiplying it by a weighting factor of 0.1, thus reducing the value of sensed chemoattractant in the illuminated region. The illumination causes exposed areas of the blob to migrate away from exposed regions. For combined $+ve$ and $-ve$ stimuli the scheduler alternated between $+ve$ and $-ve$ stimuli every 10 steps.

We represented the movement of the original vehicle movement position $p$ by an outward spiral movement initiated in the centre of the arena. We recorded tracking of the moving signal by the blob in both noise-free and noise contaminated conditions. For the noise contaminated signals the original X-axis and Y-axis position of the vehicle $p$ was recorded and each axis position was contaminated with Gaussian noise of $\sigma = 20$ to give the noisy stimulus $n$. The noisy stimulus $n$ was projected to the blob by a temporary projection of point attractant stimulus or masked repellent region. The aggregate position of the blob was recorded by the centroid of all particles comprising the blob $b$. The position of the moving stimulus was updated every 25 steps. As the distance from the start position increased, the displacement of the simulated vehicle at each movement update became larger. The positions of $p$, $n$ and $b$ were recorded along with the absolute error (in pixels) of the position of the noisy stimulus $n$ compared to original stimulus $p$ and the error of the position of the original stimulus $p$ compared to the blob centre $b$.

A typical example of the tracking behaviour of the blob under different stimuli types in noise-free conditions is shown in Fig. 6. At the start of the experiment 1500 particles comprising the blob are initialised at random positions within an $80 \times 80$ window at the centre of the arena and after approximately 250 steps the particles coalesce into an amorphous blob shape. In Fig. 6a-c the original trajectory of the moving target is indicated by the pale trace spiral and the actual path taken by the blob is indicated by the darker trace path. On initial inspection it would appear that the $+ve$ (Fig. 6a) or $\pm ve$ (Fig. 6b) stimuli give the most accurate tracking. However the plot of distance error between $p$ and $b$ indicate that the $+ve$ (attractant) stimuli results in the most error, which increases over time as the displacement of the vehicle per step increases. Although the traced blob paths in Fig. 6a and b are smoother than in Fig. 6c there is significant time lag between the current blob position and the actual target position to cause a larger overall error. The combined $\pm ve$ stimuli condition performs slightly better than $+ve$ stimuli alone but the best performance in tracking is attained by the $-ve$ (repellent only) stimulus condition (Fig. 6c).

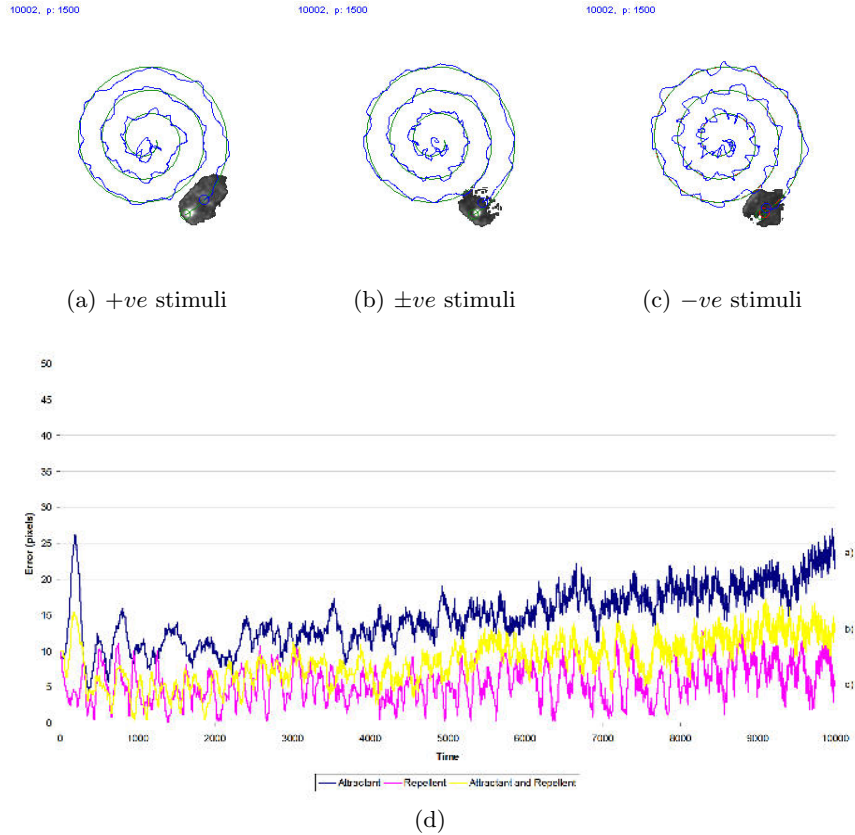

**Fig. 6.** Tracking a moving target via material adaptation and amoeboid movement under noise-free conditions. Original target path shown as light traces, blob path shown as dark traces and blob final position shown as grey mass. a) $+ve$ attractant stimuli only, b) $\pm ve$ stimuli, c) $-ve$ repellent stimuli only, d) plot of tracking error over time for methods a-c respectively.

When both the X and Y position of the moving target is contaminated with Gaussian noise of $\sigma = 20$, both the $+ve$ and $\pm ve$ conditions lose track of the target as its displacement increases (Fig. 7a and b, and the two corresponding plots in Fig. 7g). Only the $-ve$ repellent stimulus alone is able to track the target, although its error also increases over time (Fig. 7c and corresponding plot in Fig. 7g). The effect of corrupting the stimuli signal with noise in both axes is indicated in Fig. 7d-f which shows the history of noise corrupted signals superimposed on the original signal and blob position. In all three stimulus types the plots (both noise free and noise contaminated) show oscillations in the accuracy of tracking the original target. This is due to the lag time of the blob in adjusting its position (i.e. shifting its mass of particles) in response to the moving target.

## 6 Mechanisms of Material Computation

The results demonstrate that it is possible to use very simple material-based shrinkage and adaptation behaviour to indirectly reveal and summarise properties of geometric and arithmetic datasets. The actual computational mechanism is very simple: The mechanism transforms and shrinks the original shape into a minimal configuration. For a 2D shape this minimisation first withdraws external projections, then approximates a circle which, if reduced in size, will reduce to a point location. The method approximates the centroid of objects containing holes and also (although to a lesser degree of accuracy) concave objects whose centroid lies outside the shape itself. The adaptation occurs at the periphery of the mass of particles, initially withdrawing any projections into the main mass. The cohesion of the blob then pulls outward regions of the collective together and even away from the original starting shape area (in the case of concave shapes) to approximate the centroid (Fig. 3e-f and supplementary recordings). It is notable that the shrinkage of narrow projections occurs more quickly than in shapes which have a larger curve radius.

A similar effect is seen in the approximation of 1D arithmetic statistics. For an unsorted 'string' of 1D numbers the material first flattens out the peaks and troughs between the linked numbers in the material before shrinking down to a point. In the case of a sorted list of numbers the string of numbers is already almost straight and the accuracy (compared to the unsorted case) is improved because both ends of the material shrink at almost identical rates. Again, it is notable that steeper peaks and troughs are flattened out more quickly than curves with smaller gradients. Indeed, it has recently been shown that the method described in this report filters high-frequency changes in information more quickly than low-frequency changes and thus be used in data smoothing applications ([18], submitted). This data-dependent difference in behaviour may be responsible for the apparent weighting of the final position of the blob in the case of pre-sorted data with a skewed distribution.

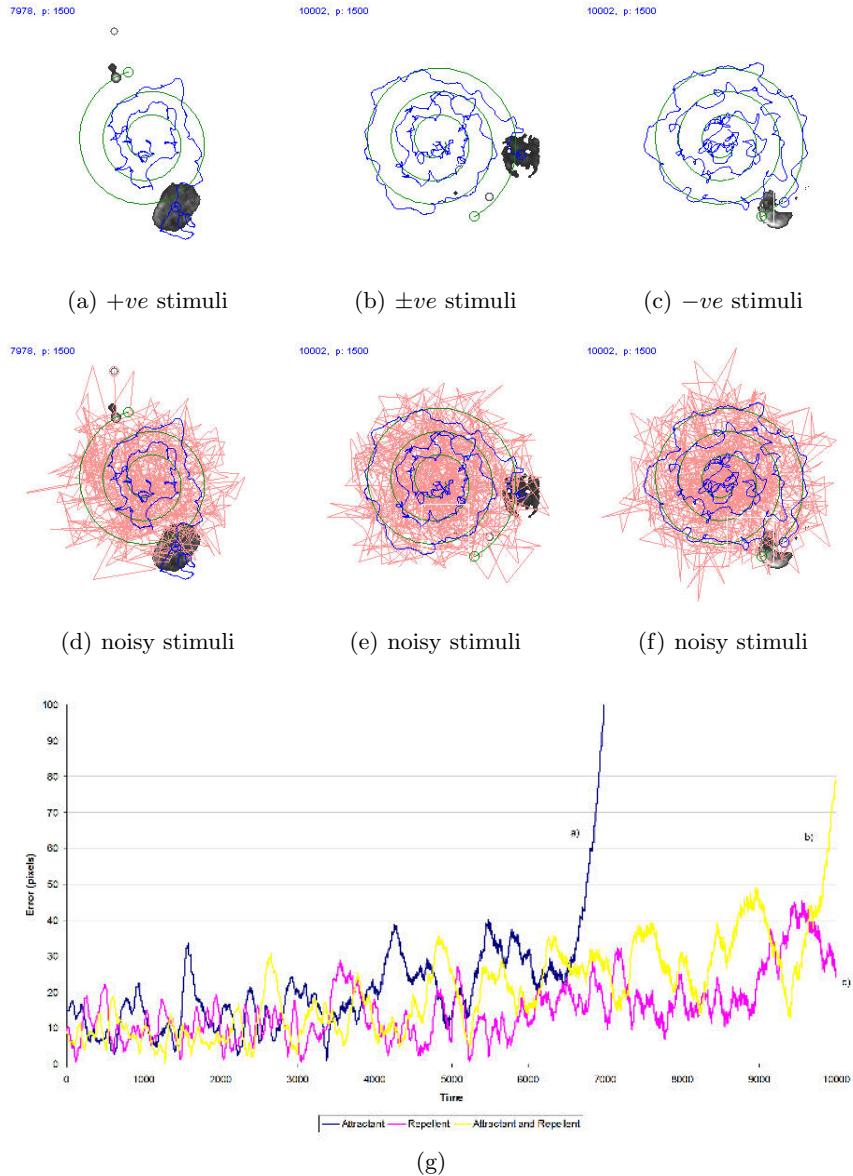

**Fig. 7.** Tracking a moving target via material adaptation and amoeboid movement under noisy stimulus conditions. Original moving target path shown as light trace, blob path tracking noisy signal shown as dark trace and blob final position shown as grey mass. a) $+ve$ attractant stimuli only, b) $\pm ve$ stimuli, c) $-ve$ repellent stimuli only, d) noise contaminated stimuli for $+ve$ condition e) noise contaminated stimuli for combined $\pm ve$ conditions, f) noise contaminated stimuli for $-ve$ condition, g) plot of tracking error over time for methods a-c respectively.

The ability of the self-oscillating amoeboid blob of material to track the noise contaminated target is, in part, due to the innate cohesion of the blob and in part due to the relatively large area of the blob. Although the blob streams towards (or away from, in the case of $-ve$ stimuli) potentially incorrect locations, the continuous update of (albeit noisy) positions ensures that the blob is not influenced too strongly by any one particular data point. Furthermore, any particularly erroneous data point stimulus actually has less of an effect on the blob than smaller errors. For $+ve$ stimuli with a very small error margin from the true location (e.g. code 1 in Fig. 8, top) this will cause regions on both sides of the blob to move towards the stimulus site. For moderately large errors (code 2 in Fig. 8, top) the blob will stream outwards towards the erroneous stimuli but the cohesion of the remainder of the blob will give a resistant force against deformation. A large erroneous stimulus (code 3 in Fig. 8, top) is less likely to influence the blob at all, since the blob is attracted more strongly to closer stimuli.

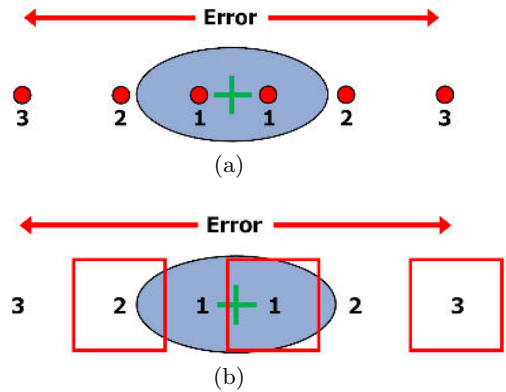

**Fig. 8.** Effect of increasingly erroneous signals on blob tracking response for attractant and repellent stimuli. Blob indicated by blue oval, actual target location indicated by green cross, erroneous $+ve$ location stimuli as red circles, erroneous $-ve$ location stimuli as red squares. Larger numbers indicate increasingly large errors of position. Top: position of erroneous $+ve$ point attractant stimuli, Bottom: position of erroneous $-ve$ stimuli mask zones.

A similar situation occurs with $-ve$ stimuli where areas outside the masking square are exposed to repellent stimuli. Small deviations from the true signal (e.g. code 1 in Fig. 8, bottom) result in areas on both sides of the square migrating inside the mask, but this region is already occupied by a large proportion of unexposed particles which prevent large shifts in blob position. Larger errors

(code 2 in Fig. 8, bottom) cause an influx into the masking zone which is resisted partly by the presence of unexposed particles inside the masking zone and partly due to the relatively large migration distance. For very large errors (code 3 in Fig. 8, bottom) the entire blob is exposed to the repellent stimuli and thus the overall position of the blob does not shift significantly. This ability to integrate and absorb a large number of noisy inputs allows the blob to track the original target despite the large noise component contaminating the signal.

The representation of numerical data in classical algorithms is efficient and inexpensive, in terms of computer resources. The transformation to a spatial unconventional computing representation requires significant spatial resources in terms of image area. It is not possible to reduce the representation of a single data point lower than a certain size because the emergent material effects are exerted only above a certain particle sensor size ($>= 3$ pixels). Furthermore, the speed of the model framework is mostly dependent on image area size (the diffusion of the environment at each scheduler step requires a convolution operation at each site) and partially dependent on population size. The time taken for the material to relax to a state where the result can be 'read' is also dependent on the size range of the data involved which also limits the practicality of the method as it currently stands. However it may be possible in future work to improve the spatial encoding of the numerical data used in the model to a more compact and efficient scheme.

## 7  Conclusions

Motivated by the complex computation performed by the single-celled organism slime mould *Physarum polycephalum*, we have demonstrated experiments that explore whether morphological adaptation in a distributed collective of very simple mobile-agent particles can be used to extract salient statistical properties of complex datasets. The mechanism is based upon the innate emergent properties of cohesion and shape relaxation within the population which behaves as a deformable virtual material. We found that by patterning the population inside a two-dimensional shape, a shrinking population can approximate the geometric centroid of a 2D dataset at the final site of shrinkage. This result appears to mimic — at least superficially — the process of sclerotium formation in slime mould in the presence of sustained adverse environmental conditions. By abstracting this idea to numerical 1D datasets (albeit represented in 2D space) we found that the final position of the shrinking population approximated the arithmetic mean of the datasets (more accurately in the case of pre-sorted datasets). Finally we demonstrated that a self-oscillatory 'blob' of the model slime mould which has previously been shown to be capable of self-organised amoeboid movement [17] can track the position of moving targets via attractant or repellent stimuli, or a fusion of both stimulus types. This tracking of the target was found to be resilient in the presence of uncertain or noisy information about the exact current position of the target.

The results in this paper demonstrate that the collective bulk mechanical properties which emerge from, and are embodied within, low-level local interactions, such as cohesion, minimisation and shape adaptation, can be harnessed to extract statistical properties of complex 1D, 2D and noisy dynamical datasets. The impetus behind this research is not to discover the 'best' method of calculating these functions, since numerical methods of doing so are already well understood and indeed more accurate. Instead the purpose is to try to understand what are the *minimal* requirements for discovering such information within these datasets. Future robotic devices will hopefully be able to access such embodied material computation to gain information about features of their environment which would otherwise require external or off-line processing. Mechanical analogies are also a useful method to explain and understand complex algorithms in computational geometry [32] and in statistics more generally [8]. We speculate that material computation methods may in future provide both practical methods for the distributed solution of statistical problems and also potentially generate novel statistical metrics based on collective material properties. More generally, this may provide a greater understanding of how simple organisms with limited computational abilities can approximate complex problems.

## Acknowledgements

This work was supported by the EU research project "Physarum Chip: Growing Computers from Slime Mould" (FP7 ICT Ref 316366)

## 8 Appendix: Particle Model Description

The multi-agent model of *Physarum* uses a population of coupled mobile particles with very simple behaviours, residing within a 2D diffusive lattice. The lattice stores particle positions and the concentration of a local diffusive factor referred to generically as chemoattractant. Particles deposit this chemoattractant factor when they move and also sense the local concentration of the chemoattractant during the sensory stage of the particle algorithm. Collective particle positions represent the global shape of the material.

### 8.1 Generation of Virtual Plasmodium Cohesion and Shape Adaptation

Particle updates occur in two distinct stages, the sensory stage and the motor stage. In the sensory stage, particles sample their local environment using three forward biased sensors whose angle from the forwards position (the sensor angle parameter, SA), and distance (sensor offset, SO) may be parametrically adjusted (Fig. 9a). The offset sensors generate local indirect coupling of sensory inputs and movement to generate the cohesion of the material. The SO distance is measured in pixels and a minimum distance of 3 pixels is required for strong local coupling to occur. For the experiments in this article we used an SO value of 9. During the sensory stage each particle changes its orientation to rotate (via the parameter rotation angle, RA) towards the strongest local source of chemoattractant (Fig. 9b). Variations in both SA and RA parameters have been shown to generate a wide range of reaction-diffusion patterns [12] and for these experiments we used SA 90 and RA 45 which results in stronger and more rapid adaptation and cohesion of the virtual material. After the sensory stage, each particle executes the motor stage and attempts to move forwards in its current orientation (an angle from 0–360 degrees) by a single pixel forwards. Each lattice site may only store a single particle and particles deposit chemoattractant into the lattice (5 units per step) only in the event of a successful forwards movement. If the next chosen site is already occupied by another particle the move is abandoned and the particle selects a new randomly chosen direction. The particles act independently and iteration of the particle population during both sensory and motor stages is performed randomly to avoid any artifacts from sequential ordering.

For the object tracking experiments we require a means of moving the position of the blob in response to the changing data. This is implemented by exploiting the self-organised amoeboid movement of the blob under oscillatory conditions. Instead of randomly selecting a new direction if a move forward is blocked, the particle increments separate internal co-ordinates until the nearest cell directly in front of the particle is free. When a cell becomes free, the particle occupies this new cell and deposits chemoattractant into the lattice at the new site. The effect of this behaviour is to remove the fluidity of the default movement of the population. The result is a surging, resistant pattern of movement. The strength of the resistance effect can be damped by a parameter $pID$ which determines the probability of a particle resetting its internal position co-ordinates, lower values providing stronger inertial movement. For these experiments we use a pID value of 0.05 which provides enough internal oscillations within the blob to generate amoeboid movement.

### 8.2 Environment and Problem Data Representation

Diffusion in the lattice was implemented at each scheduler step and at every site in the lattice via a simple mean filter of kernel size $3 \times 3$. Damping of the diffusion distance, which limits the distance of chemoattractant gradient diffusion, was achieved by multiplying the mean kernel value by 0.9 per scheduler step for centroid and arithmetic mean experiments, and by 0.93 for object tracking experiments (the oscillatory conditions require less damping to ensure blob cohesion in this case). The spatially implemented computation in the model requires that the data configuration be represented as a pattern within the 2D lattice. Data configurations for centroid approximation are loaded as greyscale image files and the shapes within these images are used as locations to initialise the particle population and to project virtual attractant into the diffusive lattice to initially confine the particle population. For the arithmetic mean experiments the data series were generated using pseudo-random number generators (generating numbers with a uniform distribution) and these numbers were spatially represented by marking the lattice pixel sites corresponding to these numbers in order to initialise the material at these locations. Neighbouring sites were linked by straight lines six pixels wide in order to provide a continuous path on which to inoculate the virtual plasmodium.

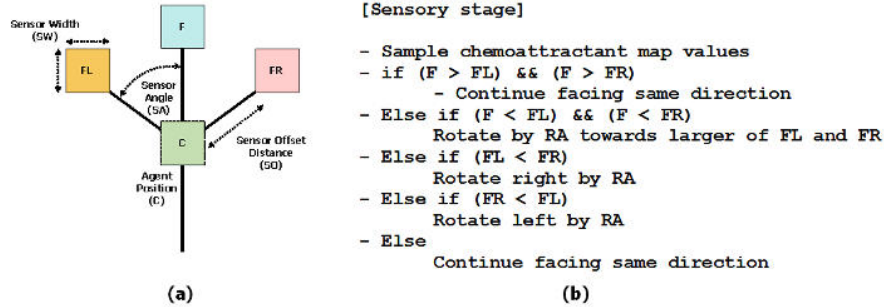

**Fig. 9.** Architecture of a single particle of the virtual material and its sensory algorithm. (a) Morphology showing agent position 'C' and offset sensor positions (FL, F, FR), (b) Algorithm for particle sensory stage.

### 8.3 Material Shrinkage Mechanism

For the centroid approximation material shrinkage was implemented by deleting a particle with probability $p = 0.0005$ at each scheduler step. When particles are removed the cohesion of the blob causes the particles to move inwards, filling the spaces and thus shrinking the blob. This simple method is sufficient when the particles comprise a large blob. For the arithmetic mean approximation, however, the thinner band of material is susceptible to breakage during shrinkage so a different method is required to ensure continuity of the material and this was implemented as follows: If there are between 1 and 10 particles (including the current particle) in a $9 \times 9$ neighbourhood of a particle, and the particle has moved forwards successfully, a new particle is created if there is a space available at a randomly selected empty location in the immediate $3 \times 3$ neighbourhood surrounding the particle. If there are between 0 and 24 particles (including the current particle) in a $5 \times 5$ neighbourhood of a particle the particle survives, otherwise it is deleted. This parameter effectively controls the speed of blob shrinkage. Deletion of a particle leaves a vacant space at this location which is filled by nearby particles, causing the collective to shrink. After a particle is deleted, the filling in of the vacant area occurs as a stochastic consequence of the particles' sensory and motor methods and no bias is introduced into the material structure. The frequency at which the growth and shrinkage of the population is executed determines a turnover rate for the particles. The frequency of testing for particle division and particle removal was every 2 scheduler steps. This relatively high frequency (compared to other applications using the virtual material approach, e.g. [20]) is due to the strong shrinkage invoked by the particular SA/RA combination used, necessitating a high adaptation frequency to maintain connectivity of the material as it adapts and shrinks.